\documentclass{iaus}
\usepackage{graphicx}

\title[Intermediate-$z$ bulges] %% give here short title %%
{The $z=0.8$ precursors of today's bulges}

\author[Balcells \& Dom\'\i nguez-Palmero]   %% give here short author list %%
{Marc Balcells \& Lilian Dom\'{\i}nguez-Palmero$^1$}

\affiliation{$^1$Instituto de Astrof\'{\i}sica de Canarias,
38200 La Laguna, Tenerife, Spain\\[\affilskip]
}

\pubyear{2008}
\volume{245}  %% insert here IAU Symposium No.
\pagerange{119--126}
\date{?? and in revised form ??}
\jname{Formation and Evolution of Galaxy Bulges}
\editors{M. Bureau, E. Athanassoula, \& B. Barbuy, eds.}
\begin{document}

\maketitle

\begin{abstract}
We study the color structure of disk galaxies in the Groth strip at redshifts $0.1<z<1.2$.  Our aim is to test formation models in which bulges form before/after the disk.   We find smooth color distributions with gentle outward blueing across the galaxy image: bulges are not distinctly redder than their disks; and bulge colors strongly correlate with global colors.  
The results suggest a roughly coeval evolution of bulges and disks.  
About 50\% of the nuclei of galaxies with central light excesses above the outer exponential profile hold passively evolving red populations.  The remainder 50\% are galaxies with central blue colors similar to their disks.  They may be bulges in formation, or the central parts of disks with non-exponential surface brightness profiles.  

%% add here a maximum of 10 keywords, to be taken form the file <Keywords.txt>
\end{abstract}

\firstsection % if your document starts with a section,
              % remove some space above using this command.
\section{The relative ages of bulges and disks}

%Bulge research is plagued with poorly posed questions about poorly defined objects.  
%This Symposium has seen strong, interesting debates on what do we understand by bulges.  Can we study objects for which we don't agree on their definition?  I think we can.  Progress in disentangling the various types of bulges, and in improving on current component decompositions of disk galaxies, requires further analysis of bulges, as well as 'may-be-bulges' and 'may-become-bulges'.

%As we learn more about the central $\sim$1 kpc of galaxies, we will have a stronger justification for adopting given terms to name specific parts of galaxies.  

%Clearly, progress is occurring, even if, often, we base our research on questions that in the end turn out to be na\"ive.  
%The work I present in this paper stems out of one such na\"ive question: 

This work addresses the simple question: 
\textsl{are bulges older than their host disks?}.   The hope is to be able to falsify one or more of the hypothesis for bulge formation: bulges before disks, via mergers or primordial collapse, \textit{vs.}\ bulges after disks, from disk instabilities or other secular processes.  
The plan is admitedly na\"ive.  A prevalence of blue bulges would argue against an old formation age for bulges.  But bulges redder than their disks would not rule out the disk instability model, as the younger disk age might come from subsequent star formation in the disk after the bulge is formed.  However, the question is a useful guideline, especially when combined with structural and isophotal diagnostics.  

For early- to intermediate-type disk galaxies in the nearby Universe, \cite[Peletier \& Balcells (1996)]{Peletier96} concluded that the age differences between bulges and disks was in most cases consistent with zero. 
But age dating of stellar populations has large uncertainties.  We have repeated the Peletier \& Balcells study at redshifts up to $z=1$, closer to the expected formation epoch of bulges.  An extensive analysis is presented in \cite[Dom\'{\i}nguez-Palmero \& Balcells (2008)]{DominguezPalmero08}.
%We define samples of disk galaxies, find the subsamples with and without bulges, and study colours measured at galactocentric distances typical of bulges.  

\section{Expectations under various bulge formation scenarios}

What should we expect under the various bulge formation scenarios? Simple-minded reasoning suggests that, 
if bulges formed before their current disks, then when looking at redshifts close to the formation epoch we should find $(a)$ bulges without disks, and $(b)$ red bulges surrounded by blue disks.  In the opposite scenario, with bulges growing out of existing disks, we should find $(a)$ disks without bulges, and $(b)$ an abundance of bulges much bluer than their parent disks (a burst would be needed to form the bulge components).  In a third scenario, where bulges and disks evolve together, we would naturally expect similar colors for bulges and disks.  Such predictions may be refined using evolutionary synthesis models.  
\cite[Bouwens, Cayon \& Silk (1999)]{Bouwens99} conclude that secular evolution models yield an overabundance of blue bulges at $z=0$ as compared to observations.  
\cite[Schulz, Fritze-v. Alvensleben \& Fricke (2003)]{Schulz03}   
show that $B/D$ appear artificially higher at high redshift, as a result of band-pass effects and evolutionary corrections.  

To the above we need to add effects that derive from the properties of the imaging survey and the adopted selection criteria. Cosmological surface brightness dimming acts against detection of extended disks, which are low surface brightness.  Hence, samples selected by flux will favor compact, spheroidal galaxies, while diameter selection will include only those bulges embedded in high-surface brightness disks.  Pixelation and PSF acts against detection of small bulges; e.g., at $z=1.0$, 1~kpc subtends 0.13 arcseconds, about one PSF of the \textsl{HST/WFPC2}.

\section{Sample and Data}

Probably, differences in sample selection account for most of the discrepancies between different studies of intermediate-$z$ bulges.    As an example, two recent conflicting claims for 'late bulge/spheroid formation' (\cite[Menanteau, Abraham \& Ellis 2001]{Menanteau01}) and for 'very red/old bulges' (\cite[Koo \etal\ 2005]{Koo05}) work with entirely different samples: visually-classified ellipsoidal galaxies in the first case, and \textsc{Gim2D} bulge-disk galaxies in the second.  

Our sample selection assumes that bulges are parts of disk galaxies, and that care should be taken to control sample pollution by bona-fide ellipticals.  Because decomposition of the surface brightness profiles carries important uncertainties, especially at high $z$, we have analyzed complementary samples with high and low axis ratios, ie. high-inclination ($50 < i < 70$ deg) and low-inclination.  The high inclination sample is almost certainly free from elliptical galaxies.  Any over-abundance of red objects in the low-inclination sample will likely be related to pollution by ellipticals.  

The results presented here refer to galaxies in the Groth-Wesphal strip (GWS).  A similar analysis has also been done in the GOODS-N field, see \cite[Dom\'{\i}nguez-Palmero \& Balcells, these proceedings]{DominguezPalmero07}.   The sample is $I$-band diameter selected (diam $> 2.8$ arcsec).  Bulge components were defined as having central brightness excesses above the outer exponential part of the surface brightness profile.  Hence, we select on density, not on the bulge protruding above the disk plane. Color profiles were measured along the semi-minor axes; the bluest side was taken to be the one freer from dust.  For both subsamples, with and without bulges, representative colors were measured at 0.85 kpc from the galaxy center.  Measuring at 0.1 arcsec\ or 0.2 arcsec hardly changes the results.  

\section{Bulge redshift distribution}

With our bulge selection criterion, we find bulges in the GSS up to $z=0.8-1.0$.  Simulations shifting galaxy images in redshift show that most galaxies fall out of the sample because the disk is lost in the noise.  Hence, bulge-disk objects such as the ones we detect might exist at higher $z$.  So do naked bulges.  However non-bulge galaxies with diameters above 2.8 arcsec are found up to $z\sim 1.2$.  And, a drop in bulge fractions was also detected in the GWS (\cite[Simard \etal\ 2002]{Simard02}) and in the HDF-N (\cite[Trujillo \& Aguerri 2004]{Trujillo04}) using flux-limited samples.

%If $z>1$ galaxies that fulfill the diameter criterion have bulge-disk structure, evolutionary and pass-band effects would tend to increase the apparent $B/D$ (\cite[Schulz, Fritze-v. Alvensleben \& Fricke 2003]{Schulz03}).  A similar drop in bulge fractions was also detected in the GWS (\cite[Simard \etal\ 2002]{Simard02}) and in the HDF-N (\cite[Trujillo \& Aguerri 2004]{Trujillo04}) using flux-limited samples.  Hence, although naked bulges might have been missed in our sample due to the diameter selection, the combined evidence suggests that bulge-disk structures truly become progressively scarce beyond $z \sim 0.8$ in the GWS.  

\begin{figure}
\begin{center}
\includegraphics[width=0.9\textwidth]{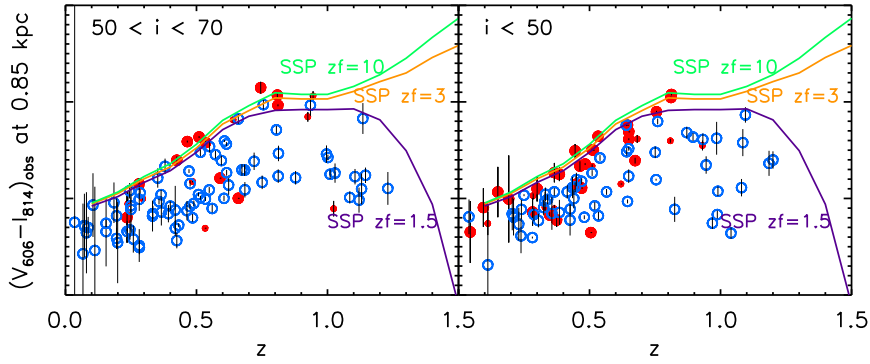}
  \caption{Observer-frame $V-I$ colors measured at 0.85 kpc from the galaxy centers, along the minor axis less affected by dust.  
    (\textit{Filled circles}) bulge sample.  
    (\textit{Open circles}) Non-bulge galaxies.  
    (\textit{a}) high-inclination sample. 
    (\textit{b}) Low-inclination sample.
    The solid lines are passive evolution tracks for the indicated formation redshifts.}\label{MB:Fig:BulgeColors}
\end{center}
\end{figure}

\begin{figure}[htbp]
\begin{center}
\includegraphics[height=4.5cm]{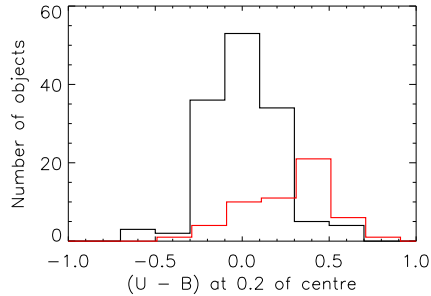}
\caption{Histogram of $U-B$ colors, measured at 0.2'' from the galaxy center, for galaxies with bulges (red) and without bulges (black).  }
\label{MB:Fig:UBhistogram}
\end{center}
\end{figure}

\section{Bulge colors}

Colors measured at 0.85 kpc from the center are shown in Figure~\ref{MB:Fig:BulgeColors}, for the high- and low-inclination samples.  For galaxies 'with bulges' (with central brightness excesses), about 50\% concentrate along the passive evolution prediction, shown by the continuous lines, while the rest show bluer colors.  Galaxies without central brightness excesses ('without bulges') scatter all over color space below the red sequence.  The extreme differences between the two rest-frame $U-B$ color distributions are clearly seen in Figure~\ref{MB:Fig:UBhistogram}.  The two samples are statistically different at above the 99.99\% level.
Bulges, as defined here, tend to populate the red sequence (RS) already at $z\sim 1$.  We concur with \cite[Koo \etal\ (2005)]{Koo05} that massive bulges tend to be very red at intermediate redshifts.  However, due to sample selection and color measurement differences, our distribution is not quite as red as that of Koo \etal, note the tail to bluer bulges.   
Because we have defined bulges on the basis of excess central surface brightness, these results suggest that stellar density might be at the origin of the migration of systems to the RS.  

Figure~\ref{MB:Fig:BulgeColors} indicates that red bulges at $0.2<z<1$ are consistent with passive evolution ($\Delta(U-B)\sim 0.2$ in two inclined bulges may reflect dust reddening). \cite[Koo \etal\ (2005)]{Koo05} found equal colors for bulges in the GWS and at $z=0$, which suggests the need for rejuvenation.  We do not find such need rejuvenation; ignoring color gradients in the disks might have slightly biased the bulge colors of Koo \etal\ to the red.  

\section{Global colors}

Analysis of global colors suggests that, instead of or in addition to stellar density, galaxy age may be a strong factor for a galaxy to populate the RS.  
%a different picture for the parameters that control bulge migration to the RS.   
When nuclear colors are compared to global colors, we find that the two strongly scale with each other.  Up to $z\sim 0.8$, redder bulges live in redder galaxies, a result that echoes a similar one found in the local Universe by \cite[Peletier \& Balcells (1996)]{Peletier96}.  We also find a scaling between bulge colors and disk colors, although the latter are subject to high measurement errors, and the relation is noisier.  On the whole, bulges are slightly redder ($\Delta(U-B) \sim 0.1-0.2$ mag) than the average disk colors.  But the bulge-disk color differential does not scale with the bulge surface brightness prominence above the exponential disk.  This indicates that bulges and disks do not show global color differences.  
%, this difference is not driven by a bluer global disk color.   
Rather galaxies show a global, gentle color gradients.  As also seen in the local Universe, the bulge-disk structure does not exist in color space.  

We failed to find red bulges surrounded by much bluer disks.  This argues against a two-phase formation of bulge-disk galaxies involving a merger origin for the bulges and a subsequent re-building of the disk, as assumed in most semi-analytic recipes, e.g., \cite[Baugh, Cole, \& Frenk (1996)]{Baugh96}.

\section{Formation scenarios}

While sample size prevents us from drawing strong conclusions, the data poses problem to the 'bulge before disk' model, given that we do not see blue disks around red bulges.  Also, bulges become progressively scarce beyond $z=0.8$.  

The 'disk before bulge' model does not fare well either.  We find few instances only of a starbursting nucleus that might correspond to a bulge in formation.  Detailed modeling of timescales needs to be carried out to constrain bulge formation rate from the observations.  

The similar colors of bulges and disks, and the global color gradient that does not show discontinuities at the bulge-disk boundary, provide some evidence for synchronized SF cessation in bulges and disks, and suggest that models in which bulges and disks follow coeval evolution, may explain most of the observations.   

\subsection{Blue 'bulges'}

As much as 50\% of the galaxies with central excess brightness ('with bulges') show nuclear colors typical of the blue cloud.  There are two basic models for these systems.  In a rejuvenation scenario, a red bulge suffers a burst of star formation due perhaps to gas accretion, and becomes momentarily blue; it will slowly redden and return to the red sequence as the burst ages (\cite[Menanteau \etal\ 2001]{Menanteau01}).  Alternatively, a blue nucleus might trace late bulge formation, or simply enhanced central star formation in the disk, perhaps related to the pseudobulge phenomenon in the local Universe.

%\section{Conclusions}\label{sec:concl}
%\begin{acknowledgments}
%\end{acknowledgments}

\end{document}